\documentstyle[12pt]{article} 
\begin{document}
\newcommand{\beq}{\begin{equation}}
\newcommand{\eeq}{\end{equation}}
\newcommand{\beqn}{\begin{eqnarray}}
\newcommand{\eeqn}{\end{eqnarray}}
\newcommand{\slp}{\raise.15ex\hbox{$/$}\kern-.57em\hbox{$\partial
$}}
\newcommand{\lnA}{\raise.15ex\hbox{$/$}\kern-.57em\hbox{$A$}}
\newcommand{\lnB}{\raise.15ex\hbox{$/$}\kern-.57em\hbox{$B$}}
\newcommand{\bP}{\bar{\Psi}}
\newcommand{\bp}{\bar{\partial}}

\begin{titlepage}

\rightline{La Plata 93/08}

\begin{center} {\large{\bf 2D Ising Model with a Defect Line}}
\end{center} \vspace{1.5cm} \begin{center} D.Cabra$^{a}$
and C.Na\'on$^{a}$ \end{center} \vspace{2cm}

ABSTRACT:{\small ~We study the two-dimensional Ising model with a defect
line and evaluate multipoint energy correlation functions
using non-perturbative field-theoretical methods. We also
discuss the evaluation of the two spin
correlator on the defect line.}

\vspace{5cm}

\noindent --------------------------------

\noindent $^a$ {\footnotesize Depto. de F\'\i sica.  Universidad
Nacional de La Plata.  CC 67, 1900 La Plata, Argentina.
Consejo Nacional de Investigaciones
Cient\'\i ficas y T\'ecnicas, Argentina.}

\end{titlepage}

\section{}

\ \indent Some time ago, Bariev \cite{B} and Mc.Coy and Perk \cite{MCP}
computed the magnetization in the (lattice) $2D$ Ising model with a defect
line (IMD). Since then, this model has remained an interesting problem in
Statistical Mechanics. One of the main features of this system is its
non-universal critical behavior, i.e. the critical exponent for the
magnetization at the defect line is a continous function of the defect strength
($\mu$) \cite{B,MCP}.
In contrast, the scaling index for the energy
density remains unchanged by the defect, as shown by Brown \cite{Br} and
Kadanoff \cite{Ka} later on.
More recently, the scaling properties
were studied by using conformal techniques \cite{HP,G,BC}. In particular,
in Ref.\cite{BC}, multipoint energy-density correlators were computed by
performing a perturbative expansion in $\mu$, in the continuum version of
the model. However, spin-spin correlators do not seem
easy to be treated within this approach.

In this letter we consider the field theoretical version of the IMD
from a non-perturbative point of view. By
extending a method first proposed by Bander and Itzykson \cite{BI} in the
context of the pure (non-defected) model, we are able to rederive the results
of Ref.\cite{BC} for the multipoint energy-density correlators in a
straightforward non-perturbative way. We discuss the application of these
ideas to the evaluation of spin-spin correlators on the line defect. We
also show how our results can be used to compute energy density correlation
functions away from the critical point.

Let us start from the generating functional of (connected) energy
density correlation functions in the massless case:

\beq
{\cal Z}_{(\mu)}[J]=\int D\bP D\Psi exp\left(-S_{(\mu)}+\int d^2z
J(z)\epsilon(z)\right) ,
\label{1}
\eeq
where:
\beq
S_{(\mu)}=\frac{1}{2\pi}\int d^2 z (\Psi \bar\partial \Psi -\bP \partial
\bP)+\mu
\int d^2 z\delta(y) \Psi \bP ,
\label{2}
\eeq
and:
\beq
\epsilon(z)=-\Psi \bP (z) ,
\label{3}
\eeq
with $\partial=\frac{\partial}{\partial z}$ $(\bp=\frac{\partial}{\bp \bar
z})$, $z=x+iy$.

The second term in (\ref{2}) arises when one considers the continuum limit
of the Ising model with a defect line \cite{B,MCP}.

In Ref.\cite{BC} Burkhardt and Choi
studied this model perturbatively, by performing an
expansion in powers of $\mu$. Using the "method of parallel
contours" these authors were able to evaluate the whole series.

We are going to rederive this result using an alternative
non-perturbative method that will be useful for the evaluation of other
relevant correlations such as those involving spin operators. To this end
we extend a technique first proposed by Bander and Itzykson
\cite{BI} in their field-theoretical calculations of the non-defected
2D Ising model spin correlator.
Following this route we are led to the computation of the fermionic Green
function associated to the action eq.(2):

\beq
\left( \begin{array}{cc}
\mu \delta (y) & \frac{1}{\pi}\partial \\
{}~~~~~~~~~~~~~~&~~~~~~~~~~~~~~~~\\
\frac{1}{\pi}\bar\partial & \mu \delta (y)
\end{array} \right)G_{\mu}(x,y;x',y')=\delta^{(2)}(x,y;x',y') .
\label{4}
\eeq

Let us introduce the complex-conjugation operator $K$, satisfying:
\beqn
K\bp K =\partial  \nonumber \\
K\partial K=\bp
\label{5}
\eeqn

As $\mu \delta(y)$ is real and concentrated in the real axis, we have simply
that $K\mu \delta (y)=\mu \delta (y) K=\mu \delta (y)$.
Using these properties, one can easily verify that the matrix operator:
\beq
{\cal C}\equiv
\frac{1}{\sqrt 2}\left( \begin{array}{cc}
K & K \\
1 & -1
\end{array} \right),
\label{6}
\eeq
diagonalizes eq.(\ref{4}).
Then the problem of calculating the Green function is reduced to the
evaluation of the diagonal elements $S_{\pm}$ of
$\tilde{G}_{\mu}\equiv {\cal C}^{-1}G_{\mu}{\cal C}$, which are determined
by the equations:

\beq
\left(\frac{1}{\pi}\bp \pm \mu \delta
(y)K\right)S_{\pm}(z,z')=\delta^{(2)}(z,z').
\label{7}
\eeq

Exactly as it happens in the pure model \cite{BI}, eq.(\ref{7}) can be easily
solved writing $S_{\pm}$ in the form:
\beq
S_{\pm}=f_{\pm}(\bar z)\frac{1}{z-z'}f_{\pm}^{-1}(\bar z'),
\label{8}
\eeq
which leads to:

\beq
\bp ln~f_{\pm}(\bar z) =\pm \pi\mu\delta(y).
\label{9}
\eeq
This equation can be readily solved and the answer is:
\beq
f_{\pm}(\bar z)=e^{\pm i\pi\mu sg(y)},
\label{10}
\eeq
where $sg(y)=\frac{y}{|y|}$. We then obtain:
\beq
S_{\pm}=\frac{e^{\pm i\pi\mu (sg~y-sg~y')}}{z-z'}.
\label{11}
\eeq

The final form for $G_{(\mu)}$ is:

\beq
G_{(\mu)}(z,z')=
\left( \begin{array}{cc}
\frac{i}{z-{\bar z}'}sin~\pi\mu (sg~y+sg~y') &
\frac{1}{z-z'}cos~\pi\mu (sg~y-sg~y') \\
{}~~~~~~~~~~~~~~~~~~~~~~~&~~~~~~~~~~~~~~~~~~~~~~\\
\frac{1}{{\bar z}-{\bar z}'}cos~\pi\mu (sg~y-sg~y') &
\frac{-i}{{\bar z}-z'}sin~\pi\mu (sg~y+sg~y')
\end{array} \right).
\label{12}
\eeq
It should be stressed that this result is exact and has not been obtained
by a perturbative expansion.
At this point the consistency of our computation can be checked by comparing
with the perurbative treatment of Ref.\cite{BC}.
To be specific, let us write down the expression for the
fermionic propagator $<\Psi (z)\bP (z')>_{(\mu)}$. Taking into account that:

\beq
G_{(\mu)}(z,z')\equiv \left<
\left(\begin{array}{c}
\Psi (z) \\
\bP (z')
\end{array}\right)
\left(\begin{array}{cc}
\Psi (z) & \bP (z')
\end{array}\right) \right>_{(\mu)},
\label{13}
\eeq
we get:
\beqn
\lefteqn{
<\Psi (z)\bP (z')>_{(\mu)}=\frac{i}{z-{\bar z}'}sin~\pi\mu (sg~y+sg~y')=}
\nonumber \\
& &\left\{\begin{array}{ll}
0, & if~y.y'<0 \\
{}~~~~~~&~~~~~~~~\\
\frac{i}{z-{\bar z}'}sin~2\pi\mu , & if~y>0~and~y'>0 \\
{}~~~~~~&~~~~~~~~\\
\frac{-i}{z-{\bar z}'}sin~2\pi\mu , & if~y<0~and~y'<0
\end{array} \right.
{}.
\label{14}
\eeqn
which exactly coincides with the result obtained in Ref.\cite{BC} after summing
thire perturbative expansion. In
particular eq.(\ref{14})
leads to the correct expresion for the v.e.v. of one energy operator.
\beq
<\epsilon(z)>_{(\mu)}=\frac{1}{|z-\bar z|}sin~2\pi\mu
\label{15}
\eeq

Let us insist that eqs.(\ref{14}) and (\ref{15}) were
obtained following an analytical method in contrast to the perturbative
approach presented in Ref.\cite{BC}.

We shall now derive from our result an explicit formula for the $n$-point
energy-density
correlation function at criticality. To this end we start from the generating
functional of connected diagrams:

\beq
log \frac{{\cal Z}_{(\mu)}[J]}{{\cal Z}_{(\mu)}[0]} =\frac{1}{2}
Tr\left\{\sum_{k=1}^{\infty}\frac{{(-1)}^{k+1}}{k}\frac{1}{2^k}
\left[\left(
\begin{array}{cc}
J&0 \\
0&J
\end{array}
\right)G_{(\mu)}\right]\right\} ,
\label{16}
\eeq
where the trace $Tr$ is defined over continuum and discrete indices
($tr_m$).

{}From eq.(\ref{16}) we obtain the explicit expression for the multipoint
correlator:
\beqn
\lefteqn{<\epsilon (z_1)...\epsilon (z_n)>_{(\mu)}=} \nonumber \\
& & \frac{(-1)^{n+1}(n-1)!}{2^{n+1}}
tr_{m}\left[G_{(\mu)}(z_1,z_2)G_{\\mu)}(z_2,z_3)...G_{(\mu)}(z_n,z_1)\right].
\label{17}
\eeqn

In particular, for $n=2$, the above formula yields:

\beqn
\lefteqn{<\epsilon (z_1)\epsilon (z_2)>_{(\mu)}=}
\nonumber \\
& & \left\{
\begin{array}{ll}
\frac{1}{|z_1-z_2|^2}-\frac{sin^22\pi\mu}{|z_1-\bar z_2|^2},  &if~~y_1y_2>0 \\
{}~~~~~~~~~~~ & ~~~~~~\\
\frac{cos^22\pi\mu}{|z_1-z_2|^2},  &if~~y_1y_2<0 .
\end{array}
\right.
\label{18}
\eeqn
which again exactly coincides with the result of Ref.\cite{BC}.

It
is certainly encouraging to correctly obtain basic features of the defected
Ising Model such as those given by eqs.(\ref{15}),(\ref{18}), by means of
a straightforward analytical procedure.

Moreover, the explicit expression obtained for the energy multipoint correlator
greatly simplifies other calculations as for example the evaluation of
correlators off the critical point, for which  a "mass term" must be added
to the free action.

Indeed, the formal expression for the connected $n$-point energy correlator
off-criticality in powers of the mass $m$ is:

\beqn
\lefteqn{<\epsilon (z_1)...\epsilon (z_n)>_{(\mu , m)}=} \nonumber \\
& & \sum_{k=0}^{\infty}\frac{m^k}{k!}\int d\omega_1 ... d\omega_k
<\epsilon (z_1)...\epsilon
(z_n)\epsilon(\omega_1)...\epsilon(\omega_k)>_{(\mu,0)}
\label{19},
\eeqn
in which the result given by eq.(\ref{17}), plays an important role. All that
rest to be done is to perform matricial traces and integrals in order to
obtain the desired perturbative order for the correlator off-criticality.

As an example, we show the explicit expression for the first order correction
to $<\epsilon(z_1)>_{(\mu, m)}$:

\beq
<\epsilon(x_1)>_{(\mu, m)}=<\epsilon(x_1)>_{(\mu, 0)}-
m\frac{\pi}{2}\frac{1}{| y_1 |} sin^2(2\pi\mu)+O(m^2).
\eeq

We conclude this letter by discussing the evaluation of the two-spin
correlator in the continuum formulation of the defected model which as far
as we know has been only computed in the lattice version.
Concerning the defect free case,
this correlator can be evaluated in the continuum version
through the formula \cite{BI}:
\beq
<\sigma_r\sigma_{r'}>_0=\left<e^{\frac{1}{2}
\int_{r}^{r'} \epsilon (x) dx}\right>_0.
\label{20}
\eeq

The generalization of this formula to the model with
a defect line has not been considered yet.

On the other hand, the exact result for the spin-spin scaling dimension
was obtained in the lattice version of the model in Ref.\cite{B,MCP} and is
given by:
\beq
\chi_{\sigma}= arccos(th~K).
\label{21}
\eeq
where the parameter $K$ is related with $\mu$ by the equation \cite{BC}:
\beq
sin~2\pi\mu= \pm th~K
\label{22}.
\eeq

In terms of $\mu$, the scaling dimension is then given by:
\beq
\chi_{\sigma}=\frac{1}{8}\pm \mu+2\mu^2
\label{23}.
\eeq

Firstly we consider the naive extension of eq.(\ref{20}) to the $\mu\neq
0$ case:

\beq
<\sigma_r\sigma_{r'}>_{(\mu)}=\left<e^{\frac{1}{2}\int_r^{r'}\epsilon (x)
dx}\right>_{(\mu)}.
\eeq

The scaling dimension obtained from this expression, after regularizing
a divergent result, is at most linear in
$\mu$.
But this is in contrast to
eq.(\ref{23}) which was found through direct lattice calculations.
This disagreement strongly suggests that the right-hand side of the above
equation is not the correct continuum limit expression for
$<\sigma_r\sigma_{r'}>_{(\mu)}$. In view of this fact we are led
to propose:

\beq
<\sigma_r\sigma_{r'}>_{\mu}=\left<e^{(\frac{1}{2}+2\mu)\int_{r}^{r'}
\epsilon (x) dx}
\right>_0,
\label{24}
\eeq
which gives the correct critical
behavior.
The validity of this expression, however, should be checked by
establishing the explicit connection between the lattice and continuum
defected models, as was done in the pure case. This is not a trivial
task and is currently under investigation.

In summary, we have presented a path-integral approach to the study of the
2d Ising model with a line defect. Our method, based on a technique first
proposed in Ref.\cite{BI}, allowed us to rederive recent results for multipoint
energy-density correlators in an elegant non-perturbative fashion. We have
also shown how to apply our procedure to multipoint correlators off
criticality.
Concerning the spin-spin correlators, we have investigated the extension
of eq.(\ref{20}), valid in the defect free case,
to the IMD.

\vspace{.5cm}

{\bf Acknowledgements}
We are grateful to G.Rossini for many fruitful conversations. We also thank
H.Falomir and E.M.Santangelo for a useful discussion and F.Schaposnik for
a careful reading of the manuscript.
D.C.Cabra and C. Na\'on are partially supported by CONICET.

\end{document}